# Research on the Detection Method of Breast Cancer Deep Convolutional Neural Network Based on Computer Aid

Mengfan Li
Xi'an Jiaotong-Liverpool University
Suzhou, China
mengfan.li20@student.xjtlu.edu.cn

*Abstract*—Traditional breast cancer image classification methods require manual extraction of features from medical images, which not only require professional medical knowledge, but also have problems such as time-consuming and labor-intensive and difficulty in extracting high-quality features. Therefore, the paper proposes a computer-based feature fusion Convolutional neural network breast cancer image classification and detection method. The paper pre-trains two convolutional neural networks with different structures, and then uses the convolutional neural network to automatically extract the characteristics of features, fuse the features extracted from the two structures, and finally use the classifier classifies the fused features. The experimental results show that the accuracy of this method in the classification of breast cancer image data sets is 89%, and the classification accuracy of breast cancer images is significantly improved compared with traditional methods.

*Keywords—computer-aided; deep learning network; deep convolutional network; breast cancer medical image; breast cancer recognition*

## I. INTRODUCTION

Breast cancer ranks first in the global incidence and mortality of female cancer. 24.2% of female cancer patients worldwide are affected by breast cancer each year, and 15% of female cancer deaths are breast cancer patients. The situation of breast cancer in China is more severe, with the incidence and mortality rate increasing every year, and the proportion of young women in the affected population is also increasing. Although the incidence of breast cancer is increasing year by year, the number of deaths due to breast cancer in developed countries such as Europe and the United States has begun to show a downward trend.

"Early detection, early treatment" is the most important way to reduce breast cancer mortality when the cause of breast cancer is uncertain. Early detection of breast cancer includes screening and diagnosis. Screening refers to the screening of people who may have breast disease. Diagnosis refers to the diagnosis of suspicious patients who have been screened out, knowing the benign and malignant areas of the lesion. Screening has high requirements for the inspection machine, while diagnosis requires a high level of imaging physicians. Mammography is the most common method for preliminary screening of breast cancer. This method is not only cheap and less painful for patients, but also provides a clear image of the structure of breast tissue [1]. Imaging physicians use breast imaging to determine whether the breast has occurred. For the benign and malignant lesions and the lesion area, the accuracy of this initial diagnosis completely depends on the prior experience of the imaging doctor, and it is easy to miss and misdiagnose. Doctors often need to evaluate benign and malignant breast tumors based on ultrasound images, and the results of the evaluation are an important prerequisite for further determining the treatment plan. However, the current diagnosis of benign and malignant tumors through ultrasound images is very dependent on the personal experience of doctors. If a relevant computer-aided diagnosis (CAD) system can be established for the auxiliary classification of breast tumors, it will improve the clinical diagnosis efficiency and reduce the misdiagnosis rate. Both have important meaning.

The classification of breast ultrasound images is a very challenging task: First, the quality of ultrasound imaging is poor, and the images often contain a lot of noise, which makes automatic classification and diagnosis more difficult; second, there is still a lack of large-scale open and labeled breast ultrasound. Image data sets bring certain difficulties to the research and testing of algorithms. At present, the common methods of medical image classification can be divided into two types, one is the classification method based on manual features, and the other is the classification method based on deep learning. In recent years, deep learning technology has made breakthrough progress in the classification of natural images. Different from traditional methods, deep learning does not need to go through artificially designed target detection, target segmentation, feature extraction and other steps. Instead, it directly inputs images and image tags, and supervised learns advanced models with complex parameters, so as to the image is predicted. At present, a series of results have been achieved in the application of deep learning in the field of medical image classification, and it has also been successfully applied in mammography and pathological slice images. Based on this, this research uses the convolutional neural network (CNN) in deep learning as the main classification method, and designs a network structure that can accept coronal and cross-section input at the same time and perform classification judgments through later fusion, and evaluates the fusion method for the impact of classification performance. In addition to image information, this research also collected the description information of each lesion, and designed a novel network structure that can integrate image information and text information, so that only one network model can be trained to complete the classification and discrimination of all information.

## II. ACQUISITION OF ULTRASOUND IMAGE OF BREAST TUMOR AND ACQUISITION OF LESION SHAPE

### A. Collection of ultrasound images of breast tumors

We cooperated with the physical examination center of related affiliated hospitals to collect 145 ultrasound image samples of breast tumors, including 71 benign lesions and 74 malignant lesions [2]. The cutting image eliminates the image information in the ultrasound image that is not related to breast tumor lesions, and uses the processed image sample as a sample library for feature extraction and classification of breast tumor ultrasound images.

### B. ROI of breast tumor ultrasound image

Region calibration After acquiring ultrasound image data, in order to extract relatively accurate image features of breast tumor lesions, doctors with rich experience in breast ultrasound calibrated the tumor lesions of each ultrasound image, and generated the ultrasound image corresponding to breast tumor lesions of interest Area image, as shown in Figure 1.

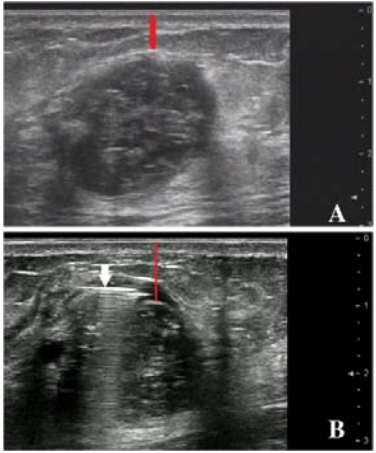

Fig. 1. The original ultrasound image of breast tumor and the ROI of the calibrated area of interest

Among them, (a) the original ultrasound image of the breast tumor, (b) the ROI of the breast tumor lesion.

### C. Obtaining the ROI shape of breast tumor lesions

Perform image median filtering, histogram equalization and other operations on the image of the region of interest of the breast tumor lesion to enhance the image. Then, Fourier transform is performed on the image, the Butterworth filter is used to eliminate the speckle noise of the ultrasound image, and the image geometric opening and closing operation is used to filter out small burrs and isolated points in the image, thereby smoothing the edge of the tumor. Perform binarization, and finally obtain the shape of the image region of interest, as shown in Figure 2.

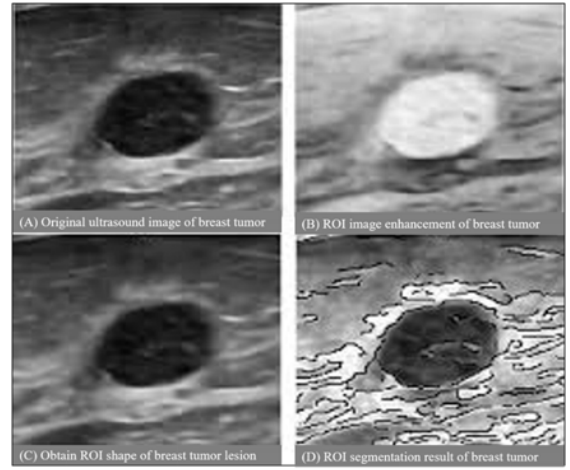

Fig. 2. Obtaining the ROI shape of breast tumor lesions

## III. CONVOLUTIONAL NEURAL NETWORK ALGORITHM

The general process of convolution operation is shown in Equation 1 [3].

$$X_j^l = f(\sum_{i \in M_j} X_i^{l-1} * w_{ij}^l + b_j^l) \quad (1)$$

In the formula, $M_j$ represents a set of input feature maps, $X_i^{l-1}$ represents the input feature maps of the $i$ convolution kernel of the $l-1$ layer network, $w_{ij}^l$ represents the $j$ weight value of the $i$ convolution kernel of the $l$ layer, and $b$ represents the bias Set the value, $f(\cdot)$ is a nonlinear activation function, * means convolution operation, and the final output $X_j^l$ is the $j$ feature map of the $l$ layer. The activation function is used to realize the nonlinear mapping of the input features of the convolutional layer, and the features in the original multi-dimensional space are mapped to another space to increase the linear separability of the data. The commonly used nonlinear activation functions of neural networks are shown in Table 1.

TABLE I. COMMON ACTIVATION FUNCTIONS

| Activation function name | Calculation formula | Formula number |
|---|---|---|
| Sigmoid function | $\sigma(x) = \dfrac{1}{1+e^{-x}}$ | (2) |
| Tanh hyperbolic tangent function | $\tanh(x) = \dfrac{e^x - e^{-x}}{e^x + e^{-x}}$ | (3) |
| ReLU modified linear unit | $f(x) = \max(0, x)$ | (4) |

The graphs of the three activation functions are shown in Figure 3. Among them, Sigmoid and tanh two functions are widely used in early neural networks, but due to their inherent shortcomings, they have been gradually replaced by ReLU functions [4]. As can be seen from the above figure, the output of the sigmoid function does not have zero-mean characteristics, which will cause the residual and accumulation of non-zero mean signals in the front-end network and cause the problem of slow convergence. In practical applications, this problem usually has a small impact on the network and can be alleviated by methods such as batch training.

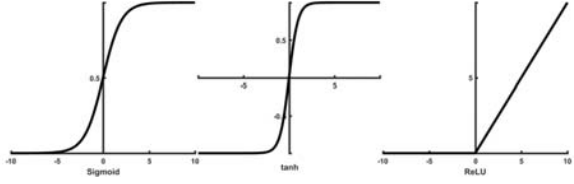

Fig. 3. Diagrams of three activation functions

In the structure of the convolutional neural network, the pooling layer is usually placed between successive convolutional layers to implement down-sampling of the feature map, and to reduce the feature map successively, thereby reducing the parameters and calculations in the network Quantity, control over-fitting. The specific pooling operation is shown in Figure 4, the filter size is 2×2, and the step size is 2.

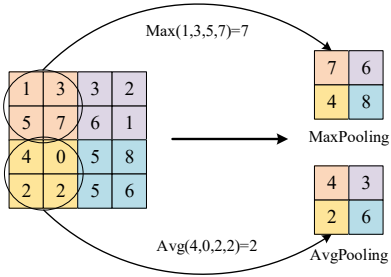

Fig. 4. Schematic diagram of pooling operation

The general calculation process of pooling can be expressed as:

$$X_j^l = f(\beta_j^l \cdot down(X_j^{l-1}) + b_j^l) \qquad (2)$$

In the formula, $\beta_j^l$ and $b_j^l$ are the multiplicative bias and additive bias of the $j$ neuron of the $l$ layer network, respectively, and F represents the sampling function. The models mentioned in this chapter all use the maximum pooling function [5]. Research on texture feature analysis in the field of computer vision shows that maximum pooling has a good effect on signal features with local sequence and periodicity. Its structure is shown in Figure 5.

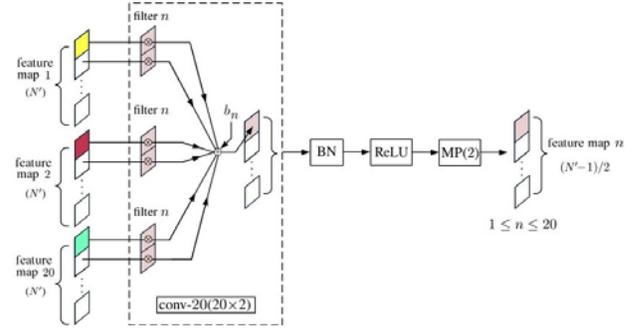

Fig. 5. Schematic diagram of fully connected layer

The most commonly used loss function in convolutional neural networks is the cross-entropy loss function, and the specific expression is shown in Equation 6.

$$Loss = H(p(x), q(x)) = -\Sigma p(x) \log q(x) \qquad (3)$$

In the formula, $p(x)$ is the target value distribution, and $q(x)$ is the estimated value distribution.

## IV. CASE ANALYSIS

### A. Experimental process

In the actual application of three-dimensional breast ultrasound, doctors will use a combination of coronal and cross-sectional reading methods for diagnosis. Three-dimensional ultrasound reconstruction of the convergent sign, irregular burr sign and complete interface echo in the coronal section is of great value for the differential diagnosis of breast tumors. At the same time, the shape and sound and shadow information on the cross section are quite different from the coronal plane [6]. According to this characteristic, this research attempts to design a new CNN network structure, which can use both the coronal and cross-sectional ROI images for classification judgment to improve the accuracy. It should be noted that the image characteristics of the coronal section and the cross section are quite different, such as the rear echo on the cross section, which is not reflected on the coronal plane. According to the principle of convolutional neural network, this study designed two independent convolutional structures for feature extraction of two sections. After the feature extraction is completed, the results of the two are connected, and the fully connected layer is used to achieve feature selection and weight optimization, thereby completing the classification judgment of the dual image input. This network structure is called 2Views-Net, and the specific structure is shown in Figure 6.

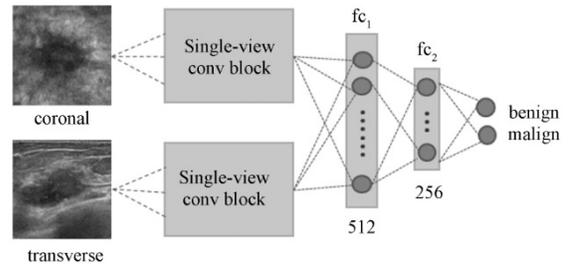

Fig. 6. 2Views-Net network structure

The input layer is to input the cross-section and coronal ROI image of the same three-dimensional data into the model. The subsequent convolution part is consistent with the structure of the Single-Net convolution part. After the convolution operation, the two output results are stretched into a one-dimensional vector, splicing and merging, and then input the fully connected layer part, $fc_1$ has a total of 512 neural nodes, and $fc_2$ has 256 neural nodes, and finally the classification is completed [7]. Compared with the fully connected layer of Single-Net, the additional nodes in the first two fully connected layers of the structure are to better adapt to the combination of cross-sectional and coronal features. At the same time, in order to clarify the fusion effect of 2Views-Net, this study also designed a probabilistic fusion method to synthesize the coronal section and cross section for classification judgment. First use the coronal image of the training set to train on Single-Net to obtain the classification model, which is recorded as Single-Net-Coronal; use the cross-sectional image of the training set to train on the Single-Net to obtain the classification model, which is recorded as Single-Net-Transverse. Assuming that the coronal image of a lesion in the test set has a predicted probability of $P_c$ obtained by the previously trained Single-Net-Coronal model, and the predicted probability of its cross-sectional image input to the Single-Net-Transverse model is $P_t$, the two Probability is fused to obtain the final probability for classification prediction. This method is referred to as Probability fusion. The process of this method is shown in Figure 7.

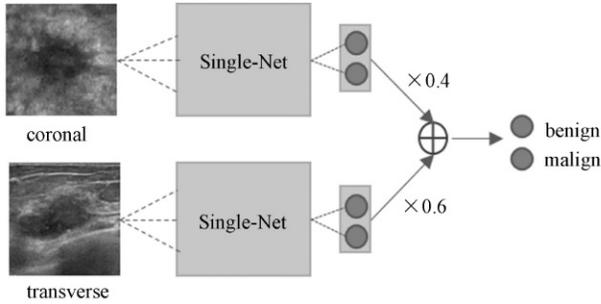

Fig. 7. Probability fusion process

*B. Experimental results*

This article selects these 3 kinds of features and the features extracted by CNN are used as the features of breast cancer images. The breast cancer image is classified, and the classification results are shown in Table 2.

TABLE II. COMPARISON OF CLASSIFICATION PERFORMANCE OF DIFFERENT FEATURES

| Method | AUC |
|---|---|
| HGD | 0.79 |
| HOG | 0.76 |
| GLCM | 0.69 |
| CNN1 | 0.84 |
| CNN2 | 0.85 |
| Feature Fusion CNN | 0.88 |

It can be seen from Table 2 that the classification performance of CNN and feature fusion CNN on the BCDR-F03 data set is better than traditional methods, reflecting the powerful feature learning ability of CNN. Among them, feature fusion CNN has the best performance, so in the following Feature fusion CNN is used in all experiments.

It can be seen from Table 3 that when the classifier uses a support vector machine, the classification performance is optimal, so in the following experiments, the support vector machine is used as the classifier.

TABLE III. CLASSIFICATION PERFORMANCE OF DIFFERENT CLASSIFIERS

| Classifier | AUC |
|---|---|
| Random forest | 0.86 |
| K neighbors | 0.85 |
| Support Vector Machines | 0.88 |

It can be seen from Table 4 that when the positive and negative samples in the training set are completely balanced, the effect is best; the more unbalanced the positive and negative samples, the worse the effect. Therefore, in this experiment, the number of benign and malignant images in the data set remains the same.

TABLE IV. THE IMPACT OF THE RATIO OF POSITIVE AND NEGATIVE SAMPLES IN THE TRAINING SET ON ACCURACY

| Positive and negative sample ratio | AUC |
|---|---|
| 2:1 | 0.86 |
| 1.3:1 | 0.88 |
| 1:1 | 0.89 |

V. CONCLUSION

This study is mainly based on the design of deep learning algorithm for tumor benign and malignant classification based on three-dimensional breast ultrasound data, and focuses on the impact of adjusting the convolutional neural network structure to integrate multiple information on classification performance. Combining the characteristics of different information and the flexibility of using the CNN model, it is proved that the use of convolutional neural network for multi-information fusion is an effective fusion method, which eliminates the steps of artificially designing fusion methods and improves classification efficiency and accuracy. This method has certain potential in solving three-dimensional data classification problems and information fusion problems, and it is worthy of further exploration and research in the future.